\documentclass[12pt,aps,floats,showpacs,amssymb,tightenlines]{revtex4}

\usepackage{amsmath}
\usepackage{amsfonts}
\usepackage{amscd}
\usepackage{epsfig}
\usepackage{amssymb}
\usepackage{tabularx}

\renewcommand{\a}{\alpha}

\newcommand{\m}{\mu}

\newcommand{\p}{\partial}

\newcommand{\s}{ Schwarzschild }
\newcommand{\x}{{x_h}}

\renewcommand{\r}{{\hat r_h}}

\begin{document}
\title{Gravitational instability of static spherically
symmetric Einstein-Gauss-Bonnet black holes in five and six
dimensions}
\author{Mart\'{\i}n Beroiz, Gustavo Dotti and Reinaldo J. Gleiser}
\affiliation{Facultad de Matem\'atica, Astronom\'{\i}a y
F\'{\i}sica (FaMAF), Universidad Nacional de C\'ordoba, Ciudad
Universitaria, (5000) C\'ordoba, Argentina}
\email{gdotti@famaf.unc.edu.ar}

\begin{abstract}
Five and six dimensional static, spherically
symmetric, asymptotically Euclidean black holes, are unstable under gravitational perturbations if
their mass is lower than a critical value set by the string tension.
The instability is due to the Gauss-Bonnet correction to Einstein's
equations, and was found in a previous
work on linear stability of Einstein-Gauss-Bonnet black holes with
constant curvature horizons in arbitrary dimensions.
We study the
unstable cases and calculate the
values of the critical masses. The results are relevant to the issue
of black hole production in high energy collisions.
\end{abstract}

\maketitle

\section{Introduction}
The most conservative approach to gravity in higher dimensions is the
one due to Lovelock \cite{Lovelock}, in which the LHS of Einstein's
equation $G_{ab} + \Lambda g_{ab} = 8 \pi G \;   T_{ab}$ is replaced with
${\cal G}_{ab}$, the most general symmetric, divergency free rank $(0,2)$ tensor
than can be constructed out of  the metric and its
first two derivatives. Lovelock's tensor is
\begin{equation} \label{l}
{\cal G}_{ab} = \sum_{n=0}^{[(D-1)/2]} c_ n G^{(n)}{}_{ab}
\end{equation}
where $D$ is the spacetime dimension, $[z]$ the highest integer
satisfying $[z] \leq z$, and $G^{(n)}{}_{ab}$ is obtained by
making appropriate contractions on a tensor product of $n$ copies of the
Riemman tensor, contractions that trivially vanish if $n > [(D-1)/2]$.

The first few $G^{(n)}{}_{ab}$'s are the spacetime metric
 ${G^{(0)}}_{ab} =
g_{ab}$, Einstein's tensor  $ {G^{(1)}}_{ab} = R_{ab}
-\frac{1}{2} R g_{ab}$, and the Gauss-Bonnet tensor
\begin{multline} \label{g2}
{G^{(2)}}_b{}^a = R_{cb}{}^{de}  R_{de}{}^{ca}  -2 R_d{}^c
R_{cb}{}^{da} -2 R_b{}^c R_c{}^a + R  R_b{}^a
 -\frac{1}{4} g^a_b \left(
R_{cd}{}^{ef}R_{ef}{}^{cd} - 4 R_c{}^d R_d{}^c + R^2 \right),
\end{multline}
If $D=4$, $G^{(n)}{}_{ab}$ vanishes for all $n>1$ and  Lovelock
theory  reduces to Einstein theory with a cosmological constant
$c_0$. Starting  with $D=5$, we may add the  ${G^{(2)}}_b{}^a$ term,
and the resulting theory, usually referred to as {\em
Einstein-Gauss-Bonnet theory} (EGB, for short), is the most general
Lovelock theory in five and six dimensions:
\begin{equation} \label{egb}
\Lambda {G_{(0)}}_b{}^a + {G_{(1)}}_b{}^a + \a {G_{(2)}}_b{}^a =
8 \pi G \; T_b{}^a,
\end{equation}
As is well known, EGB theory arises in the low energy limit of
heterotic string theories
 \cite{str1,bd}, $\alpha > 0 $ being proportional to the inverse string
tension, thus  string related treatments of BHs in higher dimension
should use the EGB equations.
 Spherically symmetric, asymptotically
Euclidean  vacuum black hole solutions of the EGB equations
(\ref{egb}) with $\Lambda=0$ are well known since the eighties
\cite{bd,w1,w2}. They are given by
\begin{equation} \label{m1}
ds^2 = -f(r) dt^2 + f(r)^{-1} dr^2 + r^2 \bar g _{ij} dx^i dx^j,
\end{equation}
$\bar g _{ij} dx^i dx^j$ the line element of $S^n$,
$n=D-2$, and
\begin{equation} \label{m2}
 f(r) = 1 +\frac{r^2}{\a(n-1)(n-2)} \left( 1 -
 \sqrt{1+\frac { 4 \a \m \left( n-1 \right)  \left( n-2 \right)}{n r^{n+1}}
  } \; \right) .
\end{equation}
$\m$ above is an integration constant, related to the mass $M$ of the black
hole through \cite{m1,m2}
\begin{equation} \label{mass}
M = \frac{\mu}{8 \pi G }  \left[ \frac{2 \pi^{\frac{n+1}{2}}}
{\Gamma \left( \frac{n+1}{2} \right) } \right] =: \frac{\mu {\cal
A}_n}{8 \pi G },
\end{equation}
${\cal A}_n$ being the area of the $n-$sphere. For positive $\mu$
and  $\alpha$, the case we are interested in, there is a single
horizon $r_h$ located at the only positive root of (note the missing
factor of $1/4$ in \cite{dg1a})
\begin{equation} \label{h1}
\mu = \frac{n r^{(n-3)}}{4}  \left[ \alpha (n-1)(n-2)+ 2 r^2
\right],
\end{equation}
then
\begin{equation} \label{masa}
M =  \frac{n r_h{}^{(n-3)} {\cal A}_n}{32 \pi G }
   \left[ \alpha (n-1)(n-2)+ 2 r_{h}{}^2
\right].
\end{equation}
The temperature and entropy of the  black hole (\ref{m1})-(\ref{m2}) are \cite{m1}
\begin{eqnarray} \label{temp}
T &=& \left[ \frac{(n-1)}{8 \pi r_h} \right]  \left( \frac{2 r_h{}^2 + \alpha (n-2)(n-3)}{r_h{}^2 + \alpha (n-1)(n-2)} \right), \\ \label{ent}
S &=& \frac{r_h{}^n {\cal A}_n}{4 G } \left[ 1 + \frac{\alpha n (n-1)}{r_h{}^2} \right]
\end{eqnarray}
The specific heat can be obtained from (\ref{masa}) and (\ref{temp}) using
\begin{equation} \label{sh1}
C = \frac{\p M }{\p T} = \left(\frac{\p M }{\p r_h}\right)
 \left(\frac{\p T}{\p r_h}\right)^{-1}.
 \end{equation}
 Introducing $\r := \frac{r_h}{\sqrt{\alpha}}$ we obtain
\begin{equation} \label{sh2}
C =  - \left( \frac{n {\cal A}_n \alpha^{n/2}}{4 G} \right) \left[ \frac{
\r ^{n-2} ( \r^2+(n-1)(n-2))^2 (2 \r^2+(n-2)(n-3))}{2 \r^4 + (n-2)(n-7) \r^2 + (n-1)(n-2)^2(n-3)} \right].
\end{equation}
Note that (\ref{m1})-(\ref{m2})  reduces to the $n+2$
dimensional Schwarzschild-Tangherlini \cite{st} black hole of
Einstein's theory in the $\a \to 0$ limit, since
\begin{equation} \label{fst}
f(r) = 1 -\frac{2 \m}{n r^{n-1}} + {\cal O} (\a).
\end{equation}
 The thermodynamic functions (\ref{temp})-(\ref{sh1}) reduce to the
 Schwarzschild-Tangherlini (ST) ones in the $\alpha \to 0$ limit.
 Note, however, that some solutions to the EGB equations  are found to
diverge as $\alpha \to 0$,
an example being the solution   (\ref{m1})-(\ref{m2})
 with a plus  sign in front of the square root
in (\ref{m2}). Other  crucial issues  strongly depend
on $\alpha$ being nonzero (we will restrict to  $\alpha >0$, as in string theory).
Consider first five dimensional ($n=3$) EGB black holes. From
(\ref{h1}) follows that there is a minimum mass $\mu = \frac{3}{2} \alpha$
for black hole formation, otherwise, (\ref{m1})-(\ref{m2}) has a naked singularity.
This does not happen for five dimensional Schwarzschild-Tangherlini (ST) black holes. The temperature
$$ T_{5D} = \frac{r_h}{2 \pi (r_h^2 + 2 \alpha)}$$
goes to infinity as $r_h \to 0^+$ ($\mu \to 0^+$) for ST holes, whereas
it tends to zero as $r_h \to 0^+$ ($\mu \to \frac{3 \alpha}{2} ^+$) in the EGB case.
The specific heat is always negative in the ST case, whereas it has a pole in the
EGB case at $r_h{} = \sqrt{2 \alpha} $, with $C > 0 $ for $ r_h < \sqrt{2 \alpha}$,
and $C < 0 $ for $ r_h > \sqrt{2 \alpha}$, i.e, small five dimensional EGB black
holes can be in equilibrium with a heat bath, contrary to what happens for ST holes.
Six dimensional EGB black holes behave more like ST black holes, their temperature decreasing monotonically from infinity in the interval $0 < r_h < \infty$, and their specific heat being always negative. However, both five and six dimensional {\em low mass} EGB black holes were found to be unstable under (linear) gravitational perturbations \cite{dg1a,dg1b,dg2}, whereas all $D>4$ ST black holes are well known to be stable under linear gravitational perturbations \cite{gh}. \\
The purpose of this work is to find the values for the
critical mass below which
five and six dimensional EGB black holes become unstable under linear gravitational perturbations. The perturbation
treatment in \cite{dg1a,dg1b,dg2} is based in the decomposition in
tensor, vector and scalar modes given in \cite{koda}, which is a
higher dimensional generalization of the axial an polar modes found in the
Regge-Wheeler treatment of Schwarzschild perturbations \cite{rw}.
The metric perturbation in the tensor modes are made from symmetric,
divergency free tensor fields $T_{ij}$ on $S^n$ satisfying $D_k D^k
T_{ij} = -k_T{}^2 T_{ij}$, $D_j$ the covariant derivative on $S^n$.
Similarly, vector (scalar)  mode perturbations are made from vector
(scalar) fields satisfying $D_k D^k T_{i} = -k_V{}^2 T_{i}$ ($D_k
D^k T = -k_S{}^2 T$). A detailed exposition of the construction of these modes
can be found in \cite{koda}. The spectrum of the Laplacian acting on
divergency free, rank $p$ symmetric tensors on $S^n$ is
\cite{higu}
\begin{equation} \label{spc}
k_p{}^2 = \ell (\ell+n-1) - p, \; \; \ell=0,1,2,...
\end{equation}
$k_S, k_V$ and $k_T$ correspond to $p=0,1$ and $2$ respectively.

\section{Scalar mode instability of five dimensional black holes}

In five dimensions there is a single horizon located at  (see (\ref{m2})-(\ref{h1}))
\begin{equation}
r_h = \sqrt{\frac{2}{3} \mu - \alpha},
\end{equation}
as long as $\mu$ is greater than $3 \alpha /2$, the minimum value
required for black hole formation. It is  convenient to adopt
the dimensionless variables from Section 4a of reference \cite{dg2}
\begin{equation} \label{dv}
x := r \alpha^{-1/2},
\;\;\; m := \frac{\mu}{\alpha},
\end{equation}
then $x_h = \sqrt{\frac{2}{3} m - 1}$ and the dimensionless tortoise
coordinate
\begin{equation} \label{star}
x^*(x) := \int_{2 \, x_h}^x \frac{dx'}{f(x')}, \hspace{.5cm} f(x) =
1 + \frac{x^2}{2} \left[ 1 - \sqrt{1 + \frac{8m}{3x}} \right]
\end{equation}
extends from minus to plus infinity.\\
Scalar perturbations in five dimensions ($n=3$) of harmonic number
$k_S{}^2 =\ell (\ell + 2), \ell = 2,3,..., $ (the modes $\ell=0,1$
are  trivial \cite{koda}) are entirely described by a single
function $\hat \phi(t,x)$ governed by an equation $\left({\cal
H}_{k_S} + \alpha \p^2/\p t^2 \right) \hat \phi =0$, which admits
separation of variables $\hat \phi(t,x) = \phi(x) e^{\omega t}$,
giving ${\cal H}_{k_S} \phi
 = -\alpha \omega^2 \phi \equiv \alpha E \phi$, with $\phi$ satisfying
 appropriate boundary conditions (reference \cite{dg2}, eqs.
(61)-(66)). The ``Hamiltonian"
\begin{equation} \label{hs}
{\cal H}_{k_S} = -\frac{\p ^2}{\p x^*{}^2} + \alpha V_{k_S}
\end{equation}
can be constructed following Section V in \cite{dg2}.
A negative eigenvalue of ${\cal H}_{k_S}$ -real $\omega$- implies
that this mode grows  exponentially with time, i.e., is unstable.
Generic perturbations have projections on each harmonic
(tensor, vector or scalar) mode. Since 5D
BHs were found to be stable under tensor and vector perturbations \cite{dg1a,dg1b,dg2}, they will  be
 unstable
  if and only if a $k_S$
is found such that the spectrum of ${\cal H}_{k_S}$ is not positive.
The boundary conditions defining the space of functions on which
${\cal H}_{k_S}$ acts determine its spectrum, $L^2(x^*,dx^*)$ being
an appropriate function space for black hole spacetimes (see
however, the discussion in \cite{dgns} regarding nakedly singular
spacetimes). The problem of stability  is then entirely equivalent
to the quantum mechanical problem of determining the sign of the
lowest eigenvalue for each member of  the  family of Hamiltonians
${\cal H}_{k_S}, k_S = \sqrt{\ell (\ell+2)}, \ell=2,3,...$
 Our strategy to prove instability
consists in showing that, if $\mu/ \alpha$ is small enough, then
for sufficiently high $k_S$, there exists
a wave function with a negative expectation value of ${\cal
H}_{k_S}$ (numerical evidence of this fact was given in Section IVa of ref \cite{dg2}). This implies that the ground state of ${\cal H}_{k_S}$
has negative energy, from where the instability
follows.\\
 From the results in Section IV of \cite{dg2}, we find,
 after a long calculation, that, after introducing
\begin{equation} x_o :=
\sqrt{\x^2+1}, \;\;\;  y := \sqrt{x^4+4 x_o{}^2},
\end{equation}
  the
potential can be conveniently split as
\begin{equation}
U_{k_S} := \frac{\alpha V_{k_S}}{f} = {k_S}^2 \; q_{\infty} + q_0 +
\frac{ {k_S}^2 q_1 + q_2}{{\cal{D}}}
\end{equation}
where ${\cal D}$ is a quartic polynomial in ${k_S}$:
\begin{equation} \label{D}
{\cal D} = 2 \; x^2\; y^4 \; \left[({k_S}^2-3) y+6 x_o{}^2) \right]^2,
\end{equation}
and  the $q'$s {\em do not depend}  on $k_S$:
\begin{eqnarray}
q_{\infty} &=& \frac{(x^4-4 x_o{}^2)}{x^2 y^2} \\
q_0 &=& \frac{(x^8 + 120 x^4 {x_o}^2- 240 {x_o}^4)}{8 x^2 y^3 } \nonumber \\
&& - \left[\frac{ x^{10} - 6 x^8 + 200 {x_o}^2 x^6 + 528 {x_o}^2 x^4
- 560 {x_o}^4 x^2 -480 {x_o}^4}{8
x^2 y^4} \right] \\
q_1 &=& 24 x^2 {x_o}^2 \left[ \left( 24 {x_o}^2 + 20 {x_o}^2 x^2 - 6
x^4-x^6 \right) y - 48 {x_o}^4 -
8 {x_o}^2x^4+x^8 \right] \\
q_2 &=& 72 x_o^2 x^4 \left[ \left(
x^6-(2x_o^2-6)x^4-20x_o^2x^2-24x_o^4-24x_o^2 \right) y \right. \nonumber \\
&&\left.  - x^8+2x_o^2 x^6+4x_o^2 x^4+40x_o^4 x^2+96x_o^4 \right]
\end{eqnarray}
Note that $q_{\infty}$ is negative in the range
$ |x| < x_c := \sqrt{2} \;
(1+\x^2)^{1/4}$, and that $0 < \x < x_c $ if and only if
 $3/2 < m < 9/2 + 3
\sqrt{2}$. Suppose this is the case and let $\psi(x)$ be a real
${\mathbb C}^{\infty}$ function vanishing outside $(\x,x_c)$,
normalized such that $$ 1 = \int_{-\infty}^{\infty} \bar \psi \psi
\; dx^* = \int_{\x}^{x_c} \frac{\psi^2}{f} dx$$ Using $\psi$ as a
test function, the expectation value of the kinetic piece of
(\ref{hs}) is
\begin{equation}
< - \p^2 / \p x^*{}^2 > = - \int_{-\infty}^{\infty} \bar \psi
\frac{\p^2 \psi}{\p x^*{}^2} \; dx^*
 = \int_{\x}^{x_c} f \left[ \frac{\p}{\p x} \left( f \frac{\p \psi}{\p x} \right) \right]^2
 dx
 \end{equation}
 and that of the scalar potential is
 \begin{equation}
  < \alpha V_{k_S} > = \alpha \int_{-\infty}^{\infty} \bar \psi V_{k_S} \psi dx^*
  = \int_{\x}^{x_c} \psi^2 \;U_{k_S} dx
  = {k_S}^2 Q_{\infty} + Q_{0} + Q({k_S})
 \end{equation}
 where
\begin{equation}
Q_{\infty}  \equiv  \int_{\x}^{x_c} \psi^2 q_{\infty} dx < 0
\end{equation}
and
\begin{equation}
 Q_{0}
\equiv  \int_{\x}^{x_c} \psi^2 q_{0} dx
\end{equation}
do not depend on $k_S$, and
\begin{equation} \label{qs}
Q({k_S}) = \int_{\x}^{x_c} \psi^2
 \left(\frac{ {k_S}^2 q_1 + q_2}{{\cal{D}}} \right) dx
\end{equation}
Note that the integrand in (\ref{qs}) converges uniformly to zero in
the interval $x \in [\x,x_c]$ as $k_S \to \infty$. This follows from
the fact that ${\cal{D}}$ is strictly positive in $[\x,x_c]$ (see
(\ref{D})) and is a quartic polynomial in $k_S$. As a consequence,
$\lim_{k_S \to \infty} Q(k_S) = 0$ and thus $ < {\cal H}_{k_S}
> $ is negative for the given test function and large values of
$k_S$. Since the above construction is possible if
\begin{equation}
3/2 < m <  m_{c(5D)} =  9/2 + 3 \sqrt{2} \simeq 8.743,
\end{equation}
we conclude that, in this mass range, all (static,
spherically symmetric, asymptotically Euclidean) 5D black holes have
a high harmonic scalar instability.\\
 Although solving the quantum mechanical
problem (\ref{hs}) analytically is out of consideration, in some
cases we were able to spot the fundamental energy using a shooting
algorithm to numerically integrate (\ref{hs}). This was done in the
standard coordinate
  $x$ (instead of $x^*$), for which (\ref{hs}) reduces to an equation of the form
  $\phi'' + P \phi ' + Q \phi =0$ with a regular singular point at the horizon.
  The first few terms of the Frobenius series around the horizon were used to generate appropriate
  initial conditions for the shooting algorithm.
  As an example, we exhibit in fig.1
 the scalar potential vs. $x$, together with the ground state wave
 function corresponding to $m = 1.7$, $\ell=2$. We also remark that
 no bound state was found for  $m > m_{c(5D)}$.

\begin{figure}[h]
\includegraphics[width=5in]{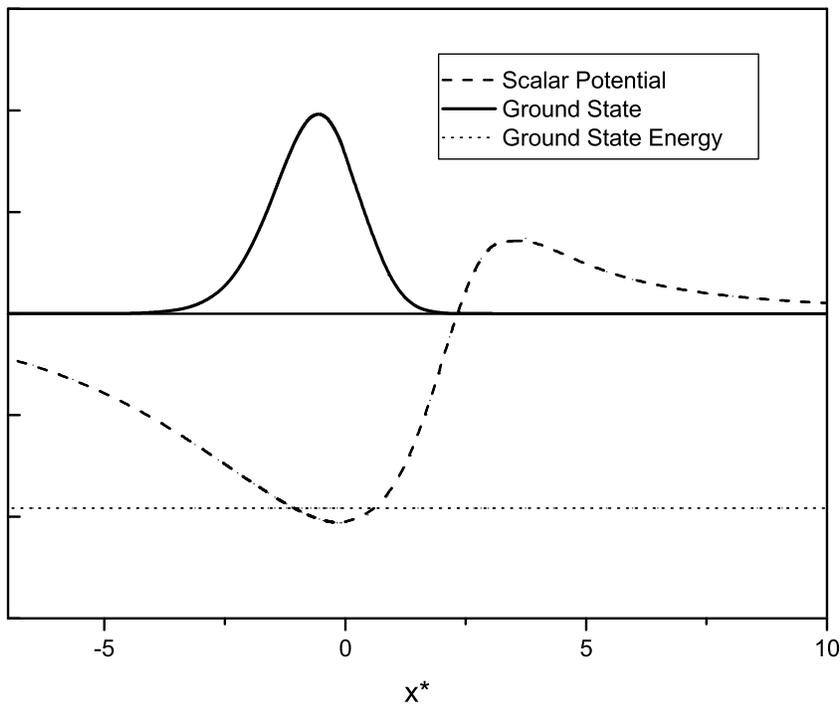}
\caption{ The scalar potential (arbitrary scale) in five dimensions
for $\ell = 2$ and $\mu / \a = 1.69$ is shown together with the
ground state wave function and energy, found numerically using a
shooting algorithm. The origin of $x^*$ was chosen as in
(\ref{star}), $x^*=0$ for $x=2 x_h$.}
\end{figure}

\section{Tensor mode instability of six dimensional black holes}

As in \cite{dg1a}, we find it convenient to introduce
dimensionless variables
\begin{equation} \label{dv2}
m := \mu \a^{-3/2}, \;\;\; x := r /(\mu \alpha)^{1/5}  = r
\alpha^{-1/2} m^{-1/5} , \;\;\; dx^* /dx := 1/f
\end{equation}
The spectrum of the Laplacian on symmetric divergency free tensors
on $S^4$ is $k_T{}^2 = \ell (\ell + 3) - 2, \ell \in {\mathbb Z}$,
only $\ell > 1$ tensors being required to construct non trivial
tensor perturbations of 6D black holes. These perturbations are
entirely described by a single function $\hat \phi(t,x)$ governed by
an equation that, after separation of variables $\hat \phi(t,x) =
\phi(x) e^{\omega t}$, assumes the form ${\cal H}_{k_T} \psi
 = -\alpha m^{2/5} \omega^2 \phi \equiv \alpha m^{2/5} E \phi$ (\cite{dg1a},
 equation (16)),
 with ``Hamiltonian"
\begin{equation} \label{ht}
{\cal H}_{k_T} = -\frac{\p ^2}{\p x^*{}^2} + \alpha
m^{\frac{2}{5}} V_{k_T},
\end{equation}
$V_{k_T}$ being the RHS of eq. (18) in \cite{dg1a}. From \cite{dg1a}
we can readily construct the potential, the result is
\begin{equation} \label{p6d1}
U := \frac{\a V_{k_T}}{f} = (k_T{}^2+2)m^{-\frac{2}{5}} \; U_o +  m^{-\frac{2}{5}}\;
 U_1 + U_2 - U_3
\end{equation}
where the $U_j$'s depend {\em only} on $x$:
\begin{eqnarray}
U_0 &=& \frac{2 (x^5 + 6)^2-75}{2  x^2(x^5+1) (x^5+6)}\\
U_1 &=& \frac{8 \;x^{20} + 72 \; x^{15} + 1218 \; x^{10} + 1752\;  x^5 -27}{4 x^2 (x^5+1)^2 (x^5+6)^2}   \\
U_2 &=&   \frac{24 \; x^{20} + 336 \; x^{15}+ 2414 \; x^{10} + 2916 \; x^5 + 189}{24 (x^5+1)^2 (x^5+6)^2}     \\
U_3 &=& \frac{ 24 \; x^{20} + 216 \; x^{15} + 1154 \; x^{10} + 1506 x^5 -81}{24 x^5 (x^5+1)^2 (x^5+6) \sqrt{1+ 6/x^5}}
\end{eqnarray}
Let $x_c = (\sqrt{75/2}-6)^{1/5} \simeq 0.658$ be the only positive
root of $U_o$, note that  $U_0 < 0$ for $0< x < x_c$. The $x$
coordinate of the horizon is
\begin{equation} \label{h}
x_h = \frac{z^2-4}{2 m^{\frac{1}{5}} \; z}, \;\; z = \left( 2 m + 2 \sqrt{16 + m^2} \right) ^{1/3}.
\end{equation}
$x_h$ is a monotone increasing function of $m$, and $x_h = x_c$ at
$m=m_{c(6D)}$ given by
\begin{equation} \label{mc6d}
m_{c(6D)} =  \frac{72 \sqrt{6} \left( 5 \sqrt{6} -12 \right)
^{3/2}}{\left( 12 - 5 \sqrt{6} + 6^{1/4} \; \sqrt{5} \sqrt{5
\sqrt{6}-12} \right) ^{5/2}}  \simeq 7.965
\end{equation}
If $m<m_{c(6D)}$, then $x_h < x_c$ and we can take a test function
supported  in $(x_h,x_c)$, so that the expectation value of the
$U_0$ piece of the potential is negative. Note from  (\ref{p6d1})
that this term is proportional to the harmonic $\ell (\ell +3)$, and
no other term of ${\cal H}_{k_T}$ depends on $\ell$, thus  the
expectation value of ${\cal H}_{k_T}$ for such a test function  will
be negative for sufficiently high harmonic number. We conclude that
6D BHs are unstable if $m < m_c$ above.\\
 Now we prove
stability for $m > m_{c(6D)}$: $U_o,U_1$ (and $U_2$) are positive if
$x>x_c$, whereas $U_2-U_3 > 0$ if  $x > x_c' \simeq 1.176$. Since
$x_h = x_c'$ for $m = m' \simeq 48.927$,  stability will follow if
we prove  that $U
> 0$ for $\ell=2,3...$,  $m_{c(6D)} < m < m'$ and $x > x_h$ given in (\ref{h}).  A lower bound
for $U$ in this region of parameter space is given by the minimum of
the single variable function $U_L := (10 U_0+U_1) 50^{-2/5} + U_2
-U_3$ in the interval $x \in (x_c,\infty)$. After some work $U_L$
can be seen to be positive in this interval, thus proving stability.
We conclude that 6D BHs are linearly unstable if and only if $ \mu /
\alpha^{3/2} =m < m_{c(6D)}$.

 Fig.2 exhibits the potential and fundamental state
(found numerically) corresponding to $\ell=2$, $\mu \a^{-3/2} =
1.85$.

\begin{figure}[h]
\includegraphics[width=5in]{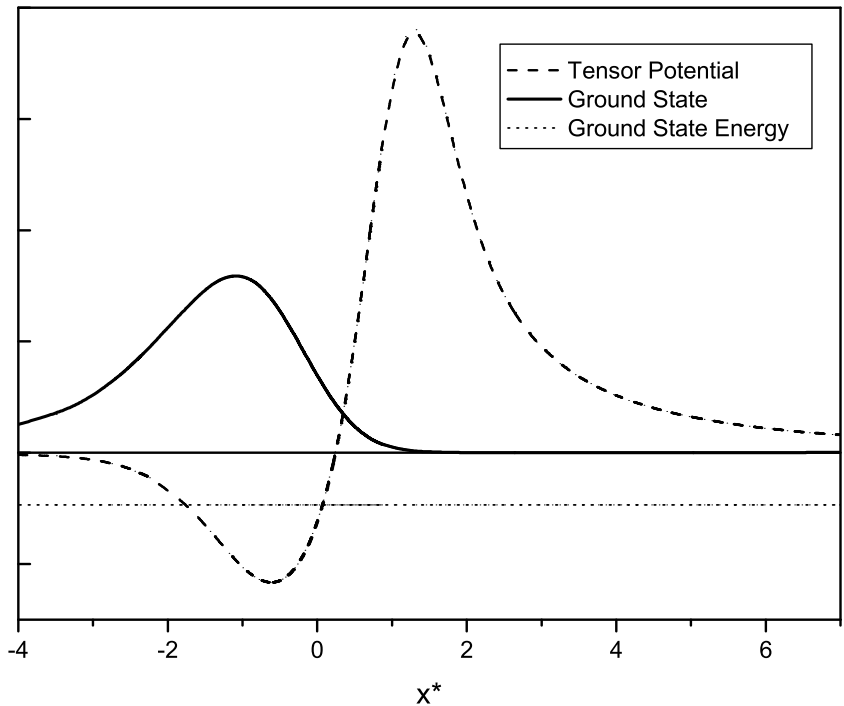}
\caption{\label{fig4} The tensor potential (arbitrary scale) in six
dimensions for $\ell = 2$ and $\mu \a^{-3/2} = 1.85$ is shown
together with the ground state wave function and energy, found
numerically using a shooting algorithm. The origin $x^*=0$
corresponds to $x=2 x_h$.}
\end{figure}

\section{Conclusions}

Gauss-Bonnet corrections to Einstein's equations in higher
dimensions have been considered in many different models, and
naturally arise in the low energy effective action of certain string
theories. However, their
 effects on black hole formation has long been disregarded. The instability
found in \cite{dg1a,dg1b,dg2} and this paper implies that the
simplest EGB black holes (asymptotically Euclidean, static,
spherically symmetric), which are the closest analogue of \s black
holes, cannot actually be formed in five space time dimensions if
their mass parameter $\mu$ (see (\ref{m2})-(\ref{mass})) is less
than $\sim 8.743 \alpha$. The Gauss-Bonnet term also prevents the
formation of these black holes in six dimensions unless $\mu$ is
greater than $\sim 7.965 \alpha^{3/2}$. The implications of these
figures depend on the context where (\ref{m1})-(\ref{m2}) is used.
As an example, the $n$-dimensional EGB black hole
(\ref{m1})-(\ref{m2}) is an approximate EGB solution if we
periodically identify one of the asymptotically Euclidean
coordinates with a period much larger than the horizon radius, and
our perturbative analysis should be valid in this setting. The large
extra dimensions scenario (suitable only for $D \geq 6$,
\cite{alexeyev}) is of interest because it allows
 $\alpha$ to be in the TeV scale \cite{alexeyev, otros},
  and so mini black holes could be produced in high energy collisions and be eventually
  detected at LHC.
  In view of our results, the probability of these events may be severely
  limited due to low mass black hole instabilities. As far as we know, this fact has not been taken into account
 in previous calculations  on black hole production rates
 in high energy collisions.
  In theories where the EGB equations simply arise as a low energy effective theory
  of some quantum gravity model,  $\alpha$
 is of the order of the Planck scale and the bounds we obtained
 for small black hole masses are much more stringent.

\section*{Acknowledgments}

This work was supported in part by grants of the Universidad
Nacional de   C\'ordoba and CONICET (Argentina).
  GD and RJG are supported by
CONICET (Argentina).

\end{document}